\documentstyle[12pt]{article}
\begin{document}
     \parindent 0pt
      \parskip 12pt
      \baselineskip 6mm

\vspace{2cm}
\begin{center}
{\Large \bf Planar random networks with flexible fibers}

\vspace{2.5cm}
\large K.J.~Niskanen$^{1}$ and M.J.~Alava$^{2}$
\\
\vspace{2cm}

\normalsize
$^{1}$KCL Paper Science Centre,
\\ P.O. Box 70, FIN-02151 Espoo, Finland\cite{kn}
\\ and
\\ USDA Forest Service, Forest Products Laboratory,
\\ One Gifford Pinchot Dr., Madison, WI 53705-2398, U.S.A.
\\
\vspace{1cm}

$^{2}$Helsinki University of Technology,
\\ Laboratory of Physics, FIN-02150 Espoo, Finland
\end{center}

\newpage
\begin{center}
{\bf Abstract}
\end{center}

The transition in random fiber networks from two-dimensional
to three-dimensional
planar structure driven by increasing coverage (total fiber
length per unit area) is studied with a deposition model.
At low coverage the
network geometry depends on the scale-free product of fiber
length and coverage while at high coverage it depends on a
scale-free combination of flexibility, width and thickness
of the fibers. With increasing coverage the roughness of the
free surface decouples from the substrate, faster when
fibers are stiffer. In the high coverage region roughness
decreases exponentially with increasing fiber flexibility.

\vspace{3cm}


\newpage

The mechanical and structural properties of two-dimensional
random systems have recently been investigated intensively
using computational and theoretical methods \cite{hr,sah}.
However, experimental studies are still few in numbers
as suitable model systems seem to be lacking.
Planar materials with random fibrous structure, such as polymer films,
short-fiber composites, paper and nonwovens, provide promising
possibilities for experimental work \cite{Kertesz}, in addition
to their intrinsic technological interest.
Fibers in real materials, such as paper, are
much longer than the sheet is thick so that the in-plane alignment of
fibers is over 90\% \cite{Rance}. The
structure is therefore planar, or "layered"
(but without well-defined layers), and the mechanical
properties are essentially those of a two-dimensional system.
The statistical geometry of strictly two-dimensional random
fiber networks is known quite well \cite{CoKa}, but
not the geometry of planar fiber networks.
Analytical models \cite{CoKa2,Gorres}
have so far not dealt properly with the bonding of
fibers that are not located in adjacent "layers". This bonding is
controlled by the random pore structure of the intervening
"layers".

In this Letter we employ a simple computer simulation
model to study how the geometry of
a random network of {\it flexible} fibers forms during the
build-up or sedimentation on a two-dimensional substrate
("growth in 2+1 D").
If coverage (i.e.~the total fiber length per unit area) is low,
the fiber networks is almost two-dimensional and the in-plane
mechanical properties are governed by coverage. At high
coverage, instead, the in-plane properties depend on
the three-dimensional "bulk" density. However, in
both cases the intrinsic property is the average number
of connections (inter-fiber bonds) per unit fiber length or
"coordination number" \cite{RBA}. This in turn
depends on the dimensions and flexibility of fibers.
We concentrate on the coverage-driven
transition from 2D to 3D network geometry with increasing
coverage and on the corresponding change in the surface
roughness \cite{Krug}.
Our results show the connection between the strictly two-dimensional
random fiber networks and their planar
three-dimensional real-world counterparts, and thereby
allow the comparison of future experiments with
theoretical or numerical studies \cite{alri,jan1,jan2}.

We start with a two-dimensional square lattice of linear
size $L = 10 \dots 1000$.
The size $L$ affects properties only close to
the percolation threshold and,
for surface roughness, at very high coverages.
Fiber length $l_f$ is an odd integer
multiple of the lattice constant ($= 1$), fiber width and thickness
are fixed, $w_f = t_f = 1$, and emphasis is given to ${l_f} \gg 1$
characteristic of fibers \cite{hf}.
Fibers are aligned at random \cite{R250} in the two principal
directions and the centers of mass are located, again at random,
over cell centers. Hence local coverage $c$ (the total
fiber length per cell) has an integer value.
Bending flexibility $T_f$ of fibers is defined
by limiting the change in the vertical elevation of a fiber
from one lattice cell to the next to $\leq T_f$ (Fig.~1).
A high $T_f$ gives a dense network, a low $T_f$ a sparse one. The
so-called wet fiber flexibility\cite{Luner} ($WFF$) of
paper-making fibers is related to $T_f$ through
$T_f = [C \times w_f \times WFF]^{1/4}$ where the constant
$C$ depends on experimental details.

During the buildup, fibers are placed down independently,
one after another, as if they were
sedimenting from a dilute suspension.
Each fiber is kept straight and parallel to the substrate until it
makes the first contact with the underlying network.
Thereafter the fiber is deformed so as to lie as
low as possible - but not below the substrate - while still
obeying the deflection constraint $T_f$. Periodic boundary conditions
are applied except in the percolation study.
The discretization of network structure should not
affect generic properties. The cumulative build-up of the
network is a more serious simplification as it does not
distinguish between densification mechanisms such as
hydrodynamic drag forces and pressing of a filtrated
fiber mat. However, the finite width
of real fibers implies that they tend
to follow a staircase pattern like that
given by the model (cf.~Fig.~1).
The construction also leads to two-sided surface properties,
but here only the free top surface is studied.

The model networks are characterized
as follows. Local thickness $t$
is the distance from the free top surface to
substrate. Thus the mean thickness ${\bar t}$
includes the voids between the
network and substrate. However, in calculating
density $\rho$ these voids are
subtracted away and so $\rho < 1$ counts only for the voids
{\it within} the sheet. This can be considered as the
"true" density, one that is independent of surface roughness.
A "coarse-grained" or "apparent"
density is also employed to mimic the finite in-plane
resolution of real measurements. In this case the
system is first divided into subgroups of 10 by 10 cells
(the size being arbitrary) and the largest value
$t = t_g$ is determined for each subgroup. The
"apparent" density is then given by $\rho_g = {\bar c}/{\bar t_g}$
where ${\bar c}$ and ${\bar t_g}$ are the mean values
of $c$ and $t_g$. Finally, the coordination
number ${\bar n}$ is defined as the average
bonded surface area of a fiber, divided by $2 l_f w_f$ \cite{RBA}.

At low mean coverage ${\bar c}$ the mechanical properties
of a planar network should be controlled by ${\bar c} - {\bar c}_c$
where ${\bar c}_c$ is the percolation threshold \cite{Kirk,Swar}
for mechanical connectivity.
We determined ${\bar c}_c$ numerically using the following rules:
1) the network is connected when there exists
a cluster of fibers that is connected
to at least one cell on two opposite boundaries of the system;
2) two cells are connected by a fiber
if the fiber covers the both sites; and 3) two fibers covering
a cell are connected, if the top surface of the lower fiber
is in contact with the bottom surface of the upper fiber.
The threshold ${\bar c}_c$
is then equivalent to the bond percolation density of the
dual lattice for bonds of length $l_f-1$. Thus an
overall ${\bar c}_c \sim 1/(l_f-1)$ behavior
is expected. On the other hand, if the network had
continuous spatial symmetry and the fibers had zero width
and infinite flexibility, then the percolation density would
be given by ${\bar c}_c = 5.71 / l_f$ \cite{PikeSeager}.

The computed percolation threshold ${\bar c}_c$
is shown in Fig.~2 for the lattice size $L = 200$.
With long fibers ($l_f \ge 20)$ the threshold is
${\bar c}_c \approx 8 /(l_f-1)$
when $T_f \rightarrow \infty$, which is slightly
larger than with continuous symmetry.
With short fibers the non-zero width
$w_f$ lowers the asymptotic value
so that ${\bar c}_c \approx 6 /(l_f-1)$ when $l_f = 5$.
The asymptotic values at $T_f \rightarrow \infty$ are
independent of $L$, increasing $L$
upto 800 revealed no deviation from the behavior shown in
Fig.~2. A random walk argument suggests that
in the opposite limit of $T_f \rightarrow 0$
the radius $r$ of a cluster with $N$ fibers is
$r \sim l_f {\sqrt N}$ and therefore the thickness
of a percolating cluster is $\sim t_f (L / l_f)^2$. Thus
the network becomes three-dimensional and ${\bar c}_c$
very large and dependent on $l_f$ and $L$.
The non-smooth small-$T_f$ behavior of ${\bar c}_c$ seen in
Fig.~2 when $l_f = 5$ is reproducible and hence
probably an artefact related to the fact that connectivity
benefits from rational values of $T_f$.
Since the maximum vertical deflection per unit length of a fiber is
$T_f$, crossover between the flexible fiber and stiff
fiber behaviors should occur at $T_f \sim t_f / l_f$;
our data suggests that this occurs at $T_f \approx 4 / l_f$.

When coverage ${\bar c}$ increases,
coordination number ${\bar n}$ increases too
and eventually becomes a constant, see Fig.~3.
In the high-coverage
region ${\bar n}$ is independent of $l_f$.
In contrast to ${\bar n}$, density $\rho$
decreases with increasing coverage
because at ${\bar c} \rightarrow 0$ there are no intra-sheet voids.
An order of magnitude higher
coverage than with ${\bar n}$ is needed before "bulk"
values of $\rho$, independent of ${\bar c}$, are obtained.
The difference is especially clear with stiff fibers. As ${\bar n}$
should be the main factor controlling the in-plane
mechanical properties of the network, it follows that
these become homogeneous, independent of coverage,
at much lower coverages than density does.

The length of a fiber $l_f$ affects the above results
at low coverages, ${\bar c} \approx {\bar c}_c$.
For example, at the percolation threshold and with large $T_f$
the number of inter-fiber bonds per
fiber, given by $2 \,{\bar n}\, l_f$ is $7.1, 6.3$ and $4.7$
for $l_f = 21, 11$ and $5$. However, at higher coverages
the three-dimensional network
structure should be controlled by the scale-free combination
$w_f T_f/t_f$. This can be rationalized by
observing that each fiber on the surface of the network prevents
new fibers from making contact with others on
$(w_f + l_f) t_f/T_f \approx l_f t_f/T_f$ sites
(on the average) when $l_f \gg w_f$, while bonding with the
fiber in question is possible in $w_f l_f$ cells.
Measured values of the wet
fiber flexibility \cite{Luner} and
known dimensions of paper-making fibers yield
$w_f T_f / t_f = 0.5 - 0.7$.

Coordination number ${\bar n}$ is tedious to measure directly
and therefore density is sometimes used to estimate it.
Indeed, Fig.~4 shows that if ${\bar c}$ is held constant
and $T_f$ is varied, then ${\bar n}$ is roughly proportional to
$\rho$ and $\rho_g$, as long as $\rho \le 0.6$
and ${\bar c}$ is not too low.
For example, in ordinary paper $\rho = 0.2 - 0.5$
so that the linear behavior should be valid.
The reproducible discontinuities at ${\bar n} \approx 0.8$ and $0.9$
(corresponding to $T_f = 2$ and $4$)
are probably again artefacts of the discrete model.
It is interesting to note that the slope
${d{\bar n}/d \rho} \approx 1.2$ with $l_f \ge 10$ and approaches
unity only for short fibers. The slope
depends slightly on coverage and thus
density usually cannot be used to compare coordination
numbers if coverage varies.

The two-dimensional projection of the network obeys
Poisson distribution \cite{CoKa} and therefore
the standard deviation of local coverage should be $\sigma_c =
{\sqrt {\bar c}}$. Our simulations follow this relationship very
accurately. The flat substrate affects the network
structure at small ${\bar c}$ but at sufficiently high ${\bar c}$
the structure is truly three-dimensional,
independent of ${\bar c}$. Then, for example,
the standard deviation of the number of pores per cell $p$ becomes
$\sigma_p = P(T_f) {\sqrt {\bar c}}$. The
amplitude $P(T_f)$ decreases with increasing $T_f$.
Areas of high coverage have few pores and therefore $P < 1$.

Simulation results for the roughness $\sigma_t$
of the free surface (i.e. the standard
deviation of $t$) can be compactly expressed as
\begin{equation}
\sigma_t^2 \approx {\bar c} \times [ A(T_f) + B({\bar c}) ]
\end{equation}
(for general $t_f$ and $w_f$ we would of course expect that
$\sigma_t \sim t_f$ and $A = A(w_f T_f / t_f)$).
The functions $A$ and $B$ are shown in Fig.~5 for
system size $L = 100$ and ${\bar c} \le 10^4$. They
are independent of fiber length as long as $l_f \ge 5$.
Equation (1) describes the coverage-driven cross-over
from a 2D to 3D network
structure. Hence, as coverage increases roughness is no longer
affected by the flat substrate. This corresponds to $B$ becoming
smaller than $A$. The decoupling is
fast with stiff fibers ($A$ is large)
because there are many pores in the network. The simulated roughness
values can easily be decomposed according to Eq.~(1) since
$\sigma_t$ becomes constant (or $A \rightarrow 0$) at large $T_f$.

The form of Eq.~(1) arose from the observation that $\sigma_t$ is
controlled by high "peaks" and deep "valleys" of the surface.
Hence the variance of $t$ could be a sum of two terms especially since
the "peaks" are dominant at low coverage or high fiber flexibility
and the "valleys" at high coverage or low flexibility.
The exponential decay of roughness at a fixed ${\bar c}$
(cf. $A(T_f)$ in Fig.~5) then follows from
the decrease in the depth of the "valleys".
The planar fiber network differs from many
growth models \cite {Krug} in that
there is nothing that would play the role of surface
tension, although fluctuations of
surface height are limited by $T_f$ over the fiber length.
Clearly, making $T_f$ small does not give a smooth surface.
In fact our model is similar to the
Vold model \cite{Vold,Krug}. The growth behavior
at high coverage, including the effects of system size and
the naturally included overhangs,
is beyond the scope of this study and will be presented elsewhere.

Summarizing, we have presented a
computer simulation study of planar random fiber networks.
The results can be compared with experiments on
disordered fibrous materials such as paper.
At high coverage ${\bar c}$ the network
is three-dimensional and its geometry is controlled by the
scale-free combination of fiber width, thickness
and flexibility, $(w_f / t_f) T_f$, provided
that $w_f / l_f \ll 1$ which usually holds fibers.
At low ${\bar c}$ the network is two-dimensional and
its properties depend on the scale-free product of coverage
and fiber length, ${\bar c}\,\,l_f$,
provided that $(l_f / t_f)\, T_f$ is large ($\geq 4$).
Thus fiber length is important only at low coverages.
Our results connect the properties of realistic planar fiber networks
with those of ideal two-dimensional networks. In particular,
if coverage is held constant then coordination number
is linearly related
to density for $\rho \le 0.6$ but the slope depends
slightly on ${\bar c}$. Roughness of the free surface decreases
exponentially with increasing $T_f$.
Because of its simplicity
our model could also be used to simulate other properties
such as three-dimensional permeability and transverse
compressibility.

This study was financially supported by the
Technology Development Centre of
Finland and by the Academy of Finland which is gratefully
acknowledged.


\newpage

\bf{Figures}

FIG. 1. Two crossing fibers on the square lattice.
The height of the "steps" is $= T_f = 1/3$.

FIG. 2. The percolation threshold $\bar{c}_c$ versus
$T_f$ for fiber lengths $l_f=$ 5, 11 and 21
(diamonds, crosses and squares, respectively);
system size $L = 200$.

FIG. 3. Coordination number ${\bar n}$,
density $\rho$ and ``apparent'' density
$\rho_g$ (triangles, squares and diamonds, respectively)
against coverage ${\bar c}$ on log-scale
with $T_f = 1$ and $2$ (filled and open symbols, respectively).
Arrows indicate the percolation point.
System size $L = 1000$, fiber length $l_f=21$.

FIG. 4. Coordination number ${\bar n}$ against density
$\rho$ and $\rho_g$ (open and closed symbols, respectively) with
growing fiber flexibility $T_f$. Coverage ${\bar c}=10.5$, $31.5$
and $94.5$ (triangles, squares and diamonds, respectively).
Fiber length $l_f=21$, system size $L=1000$.

FIG. 5. Function $A(T_f)$ on semi-log scale (Eq.~(1)), for
coverages ${\bar c} = 10, 100, 10^3$ and $10^4$ (squares,
crosses, triangles and diamonds, respectively).
Fiber length $l_f = 5$, system size $L=100$.
Inset shows function $B({\bar c})$ on semi-log scale.

\begin{thebibliography}{99}
\bibitem{hf} Fiber thickness is arbitrary in our study.
For paper-making fibers $l_f / t_f \ge l_f / w_f \simeq 20 - 50$.
\end{thebibliography}
\end{document}